\newcommand{\ud}{\rm d}
\newcommand{\un}{~\mathrm}

\newcommand{\eg}{{\em e.g. }}
\renewcommand{\vec}[1]{\mathbf{#1}}
\renewcommand{\AA}{\vec{A}}
\newcommand{\BB}{\vec{B}}
\newcommand{\CC}{\vec{C}}
\newcommand{\OO}{\vec{O}}

\documentclass[aps,prl,reprint,superscriptaddress,nofootinbib]{revtex4-1}

\usepackage{hyperref}
\usepackage[T1]{fontenc}
\usepackage{color}
\usepackage{bm}
\usepackage{amsmath} 
\usepackage{amssymb} 
\usepackage[pdftex]{graphicx}
\usepackage{xspace}
\usepackage{adjustbox}

\begin{document}

\title{Role of crystal lattice structure in predicting fracture toughness}


\author{Thuy Nguyen}
\affiliation{Service de Physique de l'Etat Condens{\'e}e, CEA, CNRS, Universit\'e Paris-Saclay, CEA Saclay 91191 Gif-sur-Yvette Cedex, France}
\affiliation{EISTI, Université Paris-Seine, Avenue du Parc, 95000 Cergy Pontoise Cedex, France}
\author{Daniel Bonamy} 
\email{Daniel.Bonamy@cea.fr}
\affiliation{Service de Physique de l'Etat Condens{\'e}e, CEA, CNRS, Universit\'e Paris-Saclay, CEA Saclay 91191 Gif-sur-Yvette Cedex, France}

\begin{abstract}
We examine the atomistic scale dependence of material's resistance-to-failure by numerical simulations and analytical analysis in electrical analogs of brittle crystals. We show that fracture toughness depends on the lattice geometry in a way incompatible with Griffith's relationship between fracture and free surface energy. Its value finds its origin in the matching between the continuum displacement field at the engineering scale, and the discrete nature of solids at the atomic scale. The generic asymptotic form taken by this field near the crack tip provides a solution for this matching, and subsequently a way to predict toughness from the atomistic parameters with application to graphene.
\end{abstract}

\maketitle

Predicting when failure occurs is central to many industrial, societal and geophysical fields. For brittle solids under tension, the problem reduces to the destabilization of a preexisting crack: Remotely applied stresses are dramatically amplified at the crack tip, and all the dissipation processes involved in the problem are confined in a small fracture process zone (FPZ) around this tip. Linear elastic fracture mechanics (LEFM) provides the framework to describe this stress concentration; the main result is that the near-tip stress field exhibits a square-root singularity fully characterized by a single parameter, the stress intensity factor, $K$ \cite{Lawn93_book,Bonamy17_crp}. Crack destabilization is governed by the balance between the flux of the mechanical energy released from the surrounding material into the FPZ  and the dissipation rate in this zone. The former is computable via elasticity and connects to $K$. The latter is quantified by the fracture energy, $\Gamma$, required to expose a new unit area of cracked surface \cite{Griffith20_ptrs}. Equivalently, it can be quantified by the fracture toughness, $K_c$, which is the $K$ value above which the crack starts  propagating. Both quantities are related via Irwin's formula \cite{Irwin57_jam}:

\begin{equation}
K_c=\sqrt{E \Gamma}
\label{eqIrwin}
\end{equation}

\noindent where $E$ is Young's modulus.

Within LEFM theory, both $\Gamma$ and $K_c$ are material constants, to be determined experimentally. Still, their value is governed by the various dissipation processes (viscous or plastic deformations, crazing, etc) at play in the FPZ. As such, it should be possible to infer $\Gamma$ and $K_c$ from the knowledge of the atomistic parameters, at least for perfectly brittle crystals in which cleavage occurs via successive bond breaking, without involving further elements of dissipation.  In Griffith's seminal work \cite{Griffith20_ptrs}, he proposed to relate $\Gamma$ to the free surface energy per unit area, $\gamma_s$, which is the bond energy times the number of bonds per unit surface: 

\begin{equation}
\Gamma=2\gamma_s
\label{eqGriffith}
\end{equation}

\noindent The factor two, here, comes from the fact that breaking one bond creates two surface atoms. This formula significantly underestimates $\Gamma$ \cite{Perez00_prl,Zhu04_prl,Buehler07_prl, Kermode15_prl}. A major hurdle is to conciliate the continuum LEFM description with the discrete nature of solids at the atomic scale: Lattice trapping effect has e.g. been invoked \cite{Thomson71_jap,Hsieh73_jap,Marder95_jmps,Marder96_pre,Bernstein03_prl,Santucci04_prl} to explain the observed shift in Griffith threshold. Analytical analysis of crack dynamics in lattices has also evidenced \cite{Slepyan81_spd,Kulakhmetova84_ms} novel high frequency waves, nonexistent in the elastic continuum, which may also explain the anomalously high value of $\Gamma$ \cite{Slepyan10_ap}. Still, till now, predicting fracture energy remain elusive.

Here, we examined how the material's resistance to crack growth is selected in two-dimensional (2D) fuse lattices of different geometries. These model digital materials, indeed, both possess perfectly prescribed and tunable "atomic bonds" and satisfy isotropic linear elasticity under antiplane deformation at the continuum scale \cite{deArcangelis89_prb,Hansen91_prl,Alava06_ap,Zapperi97_nature}. Fracture energy is observed to be significantly larger than Griffith's prediction, by a factor depending on the lattice geometry. It is the fine positioning of continuum stress/displacement fields onto the discrete lattice which sets the value of $K_c$, and consequently that of $\Gamma$. We demonstrate how the singular form of these fields near the crack tip constrains this positioning, and subsequently permits to predict fracture toughness not only in simplified antiplanar elasticity, but also in genuine elastic problems, with an application to graphene.

\begin{figure}
\begin{center}
\includegraphics[width=\columnwidth]{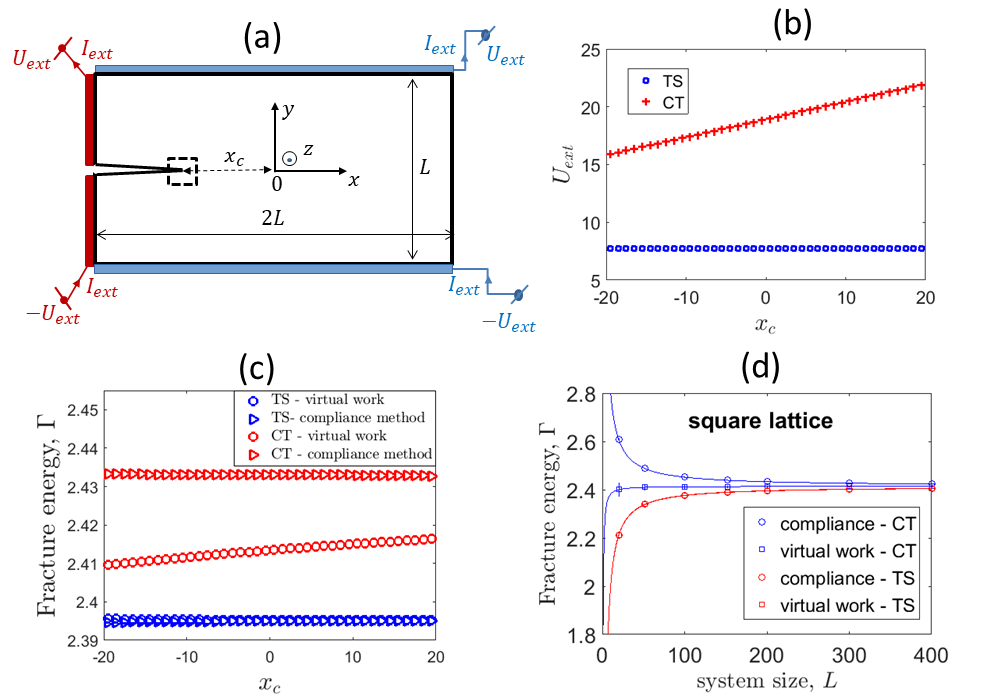}
\caption{
(a) Sketch and notation of the simulated fuse lattice. The external voltage difference $2 U_{ext}$ is imposed between the blue electrodes in the TS configuration, and between the red ones in the CT configuration. In response, a total current $I_{ext}$ flows through the network. $U_{ext}$ and $I_{ext}$ are the analog of loading displacement and loading force in experimental mechanics. 
(b) $U_{ext}$ value at breaking onset as a function of crack tip position $x_c$, in a $200\times 100$ square lattice. 
(c) $\Gamma$ vs $x_c$ in the same system. Note that $y$-axis ranges from 2.39 to 2.46. $\Gamma$ is nearly independent of $x_c$, no matter how it is measured and the loading configuration. 
(d) Size dependency of  $\Gamma$ averaged over $x_c$. For small sizes,  $\Gamma$ depends on both the measurement method and loading configuration. However, the differences vanishes as $L \rightarrow \infty$. All curves were fitted by $\Gamma=a/L^p+\Gamma_\infty$, and the fitted value $\Gamma_\infty$ is a material constant, independent of both loading configuration and measurement method. Panel (d) concerns the square geometry but the same features were observed in triangular and honeycomb lattices. Table 1 (third line) provides the values $\Gamma$ measured in the three lattice geometries.
}
\label{fig1}
\end{center}
\end{figure}

{\em Numerics --} Rectangular fuse lattices of size $2 L \times L$ in the $x$ and $y$ directions were created and a horizontal straight crack of initial length $4L/5$ is introduced from the left-handed side in the middle of the strip, by withdrawing the corresponding bonds [Fig. \ref{fig1}(a)]. This fuse network is an electrical analog of an elastic plate under antiplanar loading, where voltage field, $u(x,y)$, formally maps to out-of-plane displacement, $\vec{u}(x,y)=u(x,y)\vec{e}_z$. Triangular, square and honeycomb geometries were considered; they lead to different values for $E$ and $\gamma_s$ (Tab. 1, see Supplementary Material \cite{SuppMat} for derivation details). The length, conductance and current breakdown threshold for each fuse were assigned to unity: $\ell=1$, $g=1$ and $i_c=1$ and all quantities thereafter are given in reduced units: $\ell$ unit for length, $g$ for stresses, $i_c^2/g \ell$ for surface or fracture energy, and $i_c/\sqrt{\ell}$ for stress intensity factor or fracture toughness. 

Two different loading schemes were imposed: Thin strip (TS) configuration where the voltage difference, $2 U_{ext}$, is applied between the top and bottom nodes of the lattice, and compact tension (CT) configuration in which the voltage difference is applied between the upper (above crack) and lower (below crack) part of the left-handed side [Fig. \ref{fig1}(a)]. Voltages at each node and currents in each fuse were determined so that Kirchhoff's and Ohm's law are satisfied everywhere.  For both schemes, $U_{ext}$ was increased until the current flowing through the fuse right at the crack tip reached $i_c$; this point corresponds to the crack destabilization. This fuse was withdrawn, the crack advanced one step, and so on. The procedure was repeated until the crack has grown an additional length of $2 L/5$; at each step, the tip abscissa, $x_c$, was located at the center of mass of the next fuse about to burn.

$\Gamma(x_c)$ was computed using two procedures: (i) Virtual work method, where $\Gamma$ was computed from the loss of total energy stored in the lattice ($E_{total}=\sum_{p} i_p^2/2g$, where the current $i_p$ flows through fuse $p$) as the crack fuse is burnt  keeping $U_{ext}$ constant; (ii) Compliance method, where $\Gamma(x_c)=(1/2)(2U_{ext}(x_c))^2\ud G_{glob} /\ud x_c$ is computed from the $x_c$ variation of the global lattice conductance, ($G_{glob}=I_{ext}/2 U_{ext}$ where $I_{ext}$ is the total current flowing through the lattice, see Fig. \ref{fig1}(a)). The latter is widely used in experimental mechanics \cite{Gordon78_book} since it relates fracture energy to continuum-level scale quantities only. 

{\em Results --} Figure \ref{fig1}(b) displays examples of the profiles $U_{ext}(x_c)$ obtained in both TS and CT loading configurations: $U_{ext}$  is independent of $x_c$ in the first case, and increases with $x_c$ in the second case. Conversely, $\Gamma$ is nearly constant, within less than $0.5\%$. In-depth examination actually reveals slight dependencies with the measurement method and loading conditions [Fig. 1(c)]. The differences are significant in small systems, but they decrease with increasing $L$ [Fig. \ref{fig1}(d)]. In the continuum limit ($L \rightarrow \infty$), all values converge toward the same material constant limit, $\Gamma_\infty$, as expected in LEFM. From now on, the $\infty$ subscript is dropped for sake of simplicity. Table 1 (third line) reports the continuum-level scale values in the three lattice geometries considered here. In all cases, they are larger than Griffith's prediction (Eq. \ref{eqGriffith}), by a factor ranging between 1.75 to 3.67 depending on the geometry.

\begin{figure}
\begin{center}
\includegraphics[width=\columnwidth]{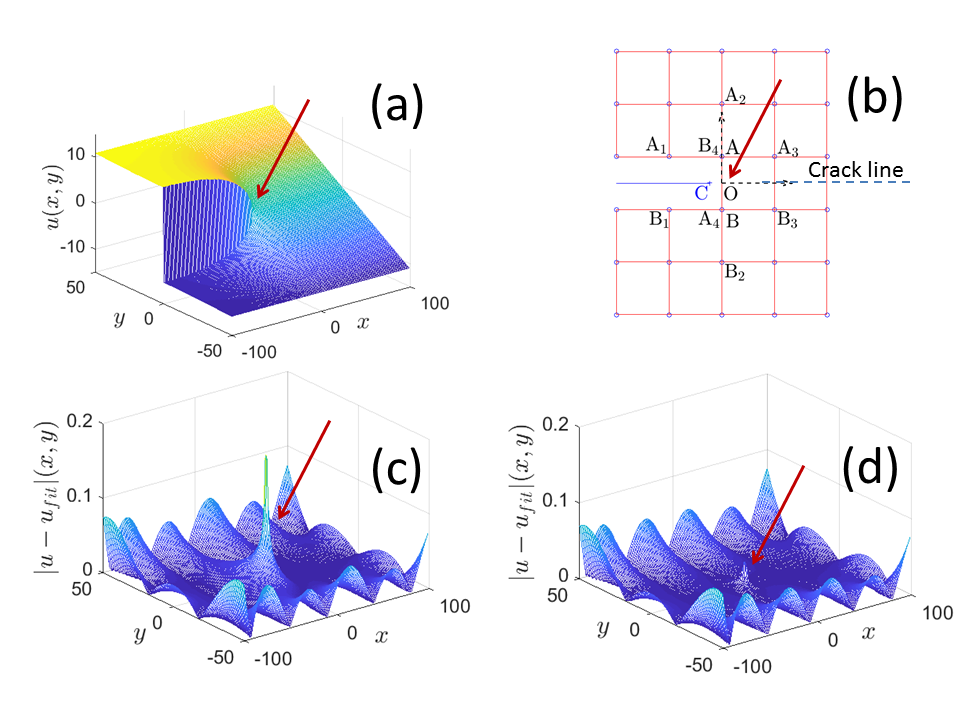}
\caption{ 
Spatial distribution of voltage field and effect of crack tip mispositioning. (a) Map of voltage field, $u(x,y)$ in a $200\times 100$ square lattice under TS loading. The crack lies along $x$ in the middle of the strip, so that the next bond about to break is at the map center (red arrow) and the external applied voltage difference, $2 U_{ext}$, is at the value required to break this bond. Equations \ref{eqWilliams} and \ref{eqWilliamsScal} are expected to describe this field. (b) Zoom on the tip vicinity emphasizing the lattice discreetness. The difficulty is to place the continuum-scale crack tip in this discrete lattice. At first, this tip is placed in the middle, $\OO$, of the bond to break. (c) Absolute difference $|u_{fit}-u|(x,y)$ between the voltage field of panel (a) and the one, $u_{fit}(x,y)$, fitted using Eqs. \ref{eqWilliams}  and \ref{eqWilliamsScal}, with $1\leq n \leq 11$ ($6$ terms). The fit is very good everywhere, except in the very vicinity of the crack tip (red arrow). (d) Absolute difference $|u_{fit}-u|(x,y)$ after correcting for the mispositioning and placing the continuum-scale tip at the proper position, $\CC$, using the iterative procedure depicted in the text. The fit is now very good everywhere.
}
\label{fig2}
\end{center}
\end{figure}

To understand the selection of $\Gamma$, we turned to the analysis of the spatial distribution of voltage [Fig. \ref{fig2}(a)]. Williams showed \cite{Williams52_jam} that, in a continuum elastic body embedding a slit crack, the displacement field writes as a series of elementary solutions $\Phi_n$:

\begin{equation}
u(r,\theta)=\sum_{n\geq 0} a_n \Phi_n(r,\theta),
\label{eqWilliams}
\end{equation}

\noindent Here, $r$ is the distance to the crack tip and $\theta$ is the angle with respect to the crack line. In the antiplane (mode III) crack problem examined here, $\Phi_n$ write (Supplementary Materials \cite{SuppMat}): 

\begin{equation}
\begin{array} {l}
\Phi^{III} _n(r,\theta) = \left\{
\begin{array}{l l}
r^{n/2}\sin{\frac{n\theta}{2}} & $if$~n~$odd$  \\
0 & $otherwise$
\end{array}
\right.
\end{array}
\label{eqWilliamsScal}
\end{equation}

\noindent The term $n=1$ is the usual square root singular term and $a_1$ relates to  $K$ via $K=E a_1\sqrt{2\pi}/4$ (see \eg \cite{Lawn93_book}). Equations \ref{eqWilliams} and \ref{eqWilliamsScal}, truncated to the first six terms ($n \leq 11$), were used to fit $u(r,\theta)$ obtained in the numerical experiments. A difficulty here is to  place properly the crack tip in the lattice; as a first guess, this tip is located in the center of the next bond to break  [Fig. \ref{fig2}(b)]. The fitted field is in good agreement with the measured one, except in the very vicinity of the crack tip [Fig. \ref{fig2}(c)]. Unfortunately, this near-tip zone is precisely the one setting whether or not the next bond breaks.    

\begin{figure}
\begin{center}
\includegraphics[width=\columnwidth]{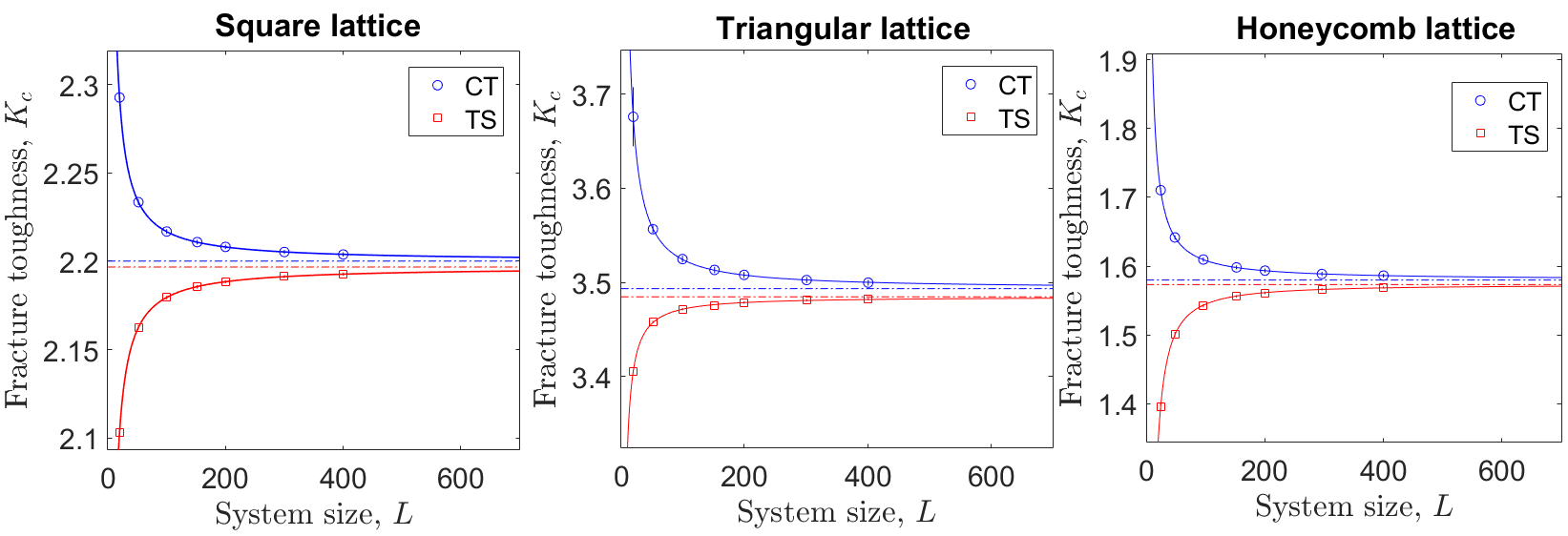}
\caption{Fracture toughness $K_c$ plotted as a function of the system size $L$, in both TS and CT loading configurations, for the three lattice geometries: square, honeycomb and triangular. $K_c$ is obtained by fitting the voltage field $u(x,y)$ using Eqs. \ref{eqWilliams} and \ref{eqWilliamsScal} after having corrected the tip mispositioning using the iterative procedure depicted in the main text. 
For all curves, the continuous line is a power-law fit $K_c = a/L^p + K_c^\infty$. The material constant $K_c^\infty$ obtained in the thermodynamic limit ($L \rightarrow \infty$) is independent of the loading conditions (dotted horizontal lines). These values are reported in the fourth line of Tab. 1.}
\label{fig3}
\end{center}
\end{figure}

The near-tip discrepancy above results from the mismatch problem between the continuum fields of LEFM and the lattice discreteness at small scale, yielding a  crack tip mispositioning. A similar problem has been faced in experimental mechanics to find out the crack tip on samples, knowing the displacement field at a set of discrete points via digital image correlation \cite{Rethore13_jmps}. This problem was solved by noting that the elementary solutions $\Phi_n$ in the Williams expansion obey \cite{Hamam07_strain,Roux06_ijf} $\partial \Phi_n/\partial x=n\Phi_{n-2}/2$. Then, a slight mispositioning of the tip position  $x$ by a small distance $d$ leads to an additional, super-singular (n=-1), term in Eq. \ref{eqWilliams}, with a prefactor $a_{-1} \simeq a_1 d /2$ \cite{Rethore13_jmps}. The crack tip can then be located as follow: (i) Start with an arbitrary mispositioned origin (in the center of the next bond to break in our case); (ii) Fit the displacement field with Eq. \ref{eqWilliams} to which the super-singular ($n=-1$) term has been added; (iii) Correct the origin position by shifting it by $d=2a_{-1}/a_1$; and (iv) Repeat this process (steps (ii) and (iii)) till the shifting distance $d$ is less than a prescribed value ($0.01\ell$ in our case). Herein, this convergence occurred in less than 3 iterations, irrespectively of the lattice geometry, lattice size and loading scheme. The obtained displacement fields fit now that measured numerically extremely well, even in the very vicinity of the crack tip [Fig. \ref{fig2}(d)]. The value of $K_c$ arises from the fitting parameter $a_1$. $K_c$ depends on $L$ for small sizes, in a way depending whether CT or TS configurations are applied. However, as $\Gamma$, $K_c$ converges toward a constant, loading independent value as $L \rightarrow \infty$ [Fig. \ref{fig3}]. The convergence is faster in lattices with higher symmetry. It also depends on loading conditions in the case with the highest symmetry (triangular lattice). Table 1 (fourth line) provides the asymptotic values obtained in the square, honeycomb and triangular geometries. In all cases, Irwin formula (Eq. \ref{eqIrwin}) relating $K_c$ and $\Gamma$ is properly satisfied, within $0.8\%$.

\begin{table}
\centering
\begin{adjustbox}{width=1\columnwidth}
\begin{tabular}{|c|c|c|c|}
\hline
Lattice geometry & Square & Triangular & Honeycomb\\
\hline
$E$ ($g$) & $2$ & $2\sqrt{3}$ & $2/\sqrt{3}$\\
\hline
$2\gamma_s$  $(i_c^2/g\ell)$ & $1$ & $2$ & $1/\sqrt{3}$\\
\hline 
$\Gamma$(sim) ($i_c^2/g\ell$) & $2.414 \pm 0.002 $ & $3.512 \pm 0.012$ & $2.152 \pm 0.012$\\
\hline 
$K_c$ (sim) ($i_c/\sqrt{\ell}$) & $2.198 \pm 0.011$ & $3.489 \pm 0.007$ & $1.576 \pm 0.008$\\
\hline
$K_c$ (approx. theo.) ($i_c/\sqrt{\ell}$) & $2.19$ & $3.55$ & $1.40$\\
\hline 
$K_c$ (exact theo.) ($i_c/\sqrt{\ell}$) & 2.24734 & 3.57137 & 1.59697\\
\hline
\end{tabular}
\end{adjustbox}
\caption{Synthesis of the different elastic and fracture parameters at play in the three lattice geometries examined here.}
\label{tab1}
\end{table}

Is it now possible to predict fracture toughness from scratch in the different lattices, without resorting to  simulations? To do so, we use the singular nature of the near-tip field and make $L \rightarrow \infty$. Consider now the very vicinity of the next bond about to break, call $\AA$ and $\BB$ the two edge atoms, and  position, as a first guess, the crack tip at the center $\OO$ of this bond [Fig. \ref{fig2}(b)]. All the $n > 1$ terms vanish in the Williams expansion and the voltage field, there, is fully characterized by two unknown only: the singular term prefactor, $a_1$, and the super-singular one, $a_{-1}$. Their value are then determined by the Kirchhoff laws at $\AA$ (or equivalently at $\BB$): $\sum_i(u(\AA_i)-u(\AA))=0$, where $\AA_i$ are the nodes themselves in contact with $\AA$. This equation gives $a_{-1}/a_1$, and subsequently an approximation of the distance $d$ between $\OO$ and the true position $\CC$ of the effective continuum-level scale crack-tip [Fig. \ref{fig2}(b)]:

\begin{equation}
d=-2\frac{\sum_i\left(\Phi^{III} _{1}(r_{\AA_i},\theta_{\AA_i})-\Phi^{III} _{1}(r_{\AA},\theta_{\AA})\right)}{\sum_i\left(\Phi^{III} _{-1}(r_{\AA_i},\theta_{\AA_i})-\Phi^{III} _{-1}(r_{\AA},\theta_{\AA})\right)}
\label{eqmisposition}
\end{equation}

\noindent Shifting the reference frame origin from $\OO$ to $\CC$ leads to the disappearance of the super-singular term. Furthermore, the prefactor of the singular term is equal to $a_1=4 K_c/E \sqrt{2\pi}$ when the force $i_{AB}=u(\BB)-u(\AA)$ applying to the bond $\AA-\BB$ is equal to the breaking threshold $i_c=1$.  This yields: 

\begin{equation}
K_c=\frac{E \sqrt{\pi}}{2\left(\Phi^{III}_1(\tilde{r}_\AA,\tilde{\theta}_\AA)-\Phi^{III}_1(\tilde{r}_\BB,\tilde{\theta}_\BB)\right)}
\label{eqKcscalar}
\end{equation}

\noindent where $(\tilde{r},\tilde{\theta})$ are the polar coordinate of the considered point in the new reference frame positioned at $\CC$. Equations \ref{eqmisposition} and \ref{eqKcscalar} provide an approximation for $K_c$ in the three lattice geometries considered here. The values are reported in Tab. 1 (fifth line) and coincide with that measured on the numerical experiments within $1\%$, $11\%$, and $2\%$  for square, honeycomb and triangular lattices, respectively. 

Note that the mispositioning $d$ obtained by Eq. \ref{eqmisposition} is an approximation only since it involves Kirchhoff's law at a single point, $\AA$, and a single iteration in the shifting procedure. Actually, both assumptions can be released by imposing Kirchhoff laws to an arbitrary number of nodes and by applying more iterations. These refinements are described in Supplementary Material \cite{SuppMat}. They provides ab-initio predictions for the fracture toughness, up to any prescribed accuracy! Table 1 (sixth line) presents the obtained values. They coincide with the numerically measured ones within $2\%$, $1.3\%$, and $2\%$ for square, honeycomb and triangular lattices, respectively.

The analytical procedure derived here for model electric/scalar elastic crack problems can be extended to genuine 2D elastic  (plane stress) crack problems. This comes about via several correspondences: (i) In such cases, the (vectorial) displacement field $\vec{u}(x,y)$ also follows Williams's asymptotic form [Eq. \ref{eqWilliams}]; and (ii) the relation $\partial \vec{\Phi}^{I}_n/\partial x=n\vec{\Phi}_n^{I}/2$ remains valid \cite{Rethore13_jmps}. Here, superscript $I$ indicates  opening fracture mode (mode I). Nevertheless, 2D crystals include an additional complicating element with respect to the fuse networks considered till now: the deformation are accommodated at the atomistic scale by two distinct modes: bond stretching and bond bending, characterized by the bond stretching stiffness $k_s$ and bond bending stiffness $k_b$ [both can be measured experimentally via infrared spectroscopy \cite{Smith98_book}]. The procedure then follows the same route as for the electrical network; it is detailed in Supplementary Material \cite{SuppMat}. It quantitatively relates $K_c$ to $E$, $k_s$, $k_b$, $\ell$, the bond strength $F_c$, the Poisson ratio $\nu$ and the crystal geometry. This procedure has been applied in graphene, which has a planar honeycomb geometry with $\ell=0.142\un{nm}$ and $E=340\un{N.m}^{-1}$ \cite{Lee08_science}, $\nu=0.18$ \cite{Liu07_prb}, $k_s\simeq 688-740\un{N.m}^{-1}$ \cite{Medina15_mp}, $k_b= 0.769 - 0.776 \times 10^{-18} \un{N\,m\,rad}^{-2}$\cite{Medina15_mp}, and $F_c=13.6\un{nN}$ \cite{Xu11_book}. It yields $K_c = 1.25\times 10^{-3}\un{N.m}^{-1/2}$ (Supplementary Material \cite{SuppMat}). Fracture energy is subsequently deduced from Irwin formula  [Eq. \ref{eqIrwin}]: $\Gamma = 4.6 \times 10^{-9} \un{J.m}^{-1}$. The predicted values for $K_c$ and $\Gamma$ are very close to the experimental ones \cite{Zhang14_natcom}: $K_c=1.4 \times 10^{-3} \un{N.m}^{-1/2}$ and $\Gamma=5.4 \times 10^{-9} \un{J.m}^{-1}$. Note that Griffith's seminal prediction [Eq. \ref{eqGriffith}] would have yielded $\Gamma = (1/2\sqrt{3})(F_c^2/k_s \ell) \simeq 5.08-5.47 \times 10^{-10} \un{J.m}^{-1}$; hence it underestimates the measured value by a factor of ten!  

{\em Concluding discussion --} This work demonstrates the key role played by the crystal lattice geometry at the atomistic scale in the selection of the resistance to failure at the macroscopic scale. It challenges Griffith's seminal interpretation \cite{Griffith20_ptrs} directly linking $\Gamma$ to $\gamma_s$. It also questions the argument proposed in \cite{Slepyan81_spd,Kulakhmetova84_ms,Slepyan10_ap} to explain Griffith's failure: Slepyan et al examined the structure of the waves emitted when cracks propagate in discrete lattices at constant speed, $v$ and, from this, they determined the fracture energy and its variation with $v$, $\Gamma(v)$. They observed that in the limit of vanishing speed, $\Gamma(v\rightarrow 0^+)$ is larger than $2\gamma_s$. They attributed the difference to the energy of phonons ( lattice-induced high-frequency waves) that would survive when $v=0$. Inertia is absent in our simulations and, hence, phonons cannot be activated. Still, we observe the same anomalously high value for fracture energy, as that obtained in dynamically cracking lattices at vanishing speeds; it is worth mentioning that the value 2.414 determined here for a square lattice [Tab. 1] coincides exactly with that reported in \cite{Slepyan81_spd}, in the $v\rightarrow 0$ limit. This means that the fracture energy in excess does not find its origin in the presence of phonons as argued in \cite{Slepyan81_spd,Kulakhmetova84_ms,Slepyan10_ap}. The interpretation we propose is that fracture toughness is fixed by the fine positioning of the near tip singularity of the continuum stress field on the crystal lattice, then by its adjustment so that the force exerted on a chemical bond is sufficient to break it. Then {\em in a second step}, fracture energy is deduced from Irwin formula [Eq. \ref{eqIrwin}].

Quantitative predictions for fracture toughness have been reached here for bidimensional crystals. The analysis is expendable to 3D and non-linear cases. This necessitates  to use the proper (and more complicated) asymptotic near-tip displacement fields; 
\eg those obtained in anisotropic elasticity for 3D crystals, the asymptotic HRR solutions \cite{Hutchinson68_jmps,Rice68_jmps} for power-law hardening solids, or those obtained in weakly non-linear elasticity, in \cite{Bouchbinder09_jmps}. The analysis can also be instrumental to explain how fracture toughness evolves with relative density and/or stockiness in cellular materials of various geometries \cite{Huang91_amsm,Fleck07_jmps,Quintana-Alonso09_chapter,Quintana-Alonso10_am}. Finally, it may catalyze further research toward novel tough, lighweight metamaterials. In this context, the use of microlattices made from periodically arranged hollow microtubes seems extremely promising \cite{Schaedler11_science,Jang13_natmat, Zheng14_science}; they both exhibit ultra-low densities and good resistance to compressive fracture. Our work provides the first step to address tensile fracture in this novel class of metamaterials. Beyond solid fracture, the analysis presented here may aid in the progression and understanding of a variety of singularity-driven systems, including the selection of the slip length in the wetting contact line problem \cite{Huh71_jcis} and the selection of the dislocation core size in crystal plasticity.

\begin{acknowledgments}
We thank Francois Daviaud, Francois Ladieu and Cindy Rountree for the careful reading of the manuscript. Funding through ANR project MEPHYSTAR (ANR-09-SYSC-006-01) and by "Investissements d'Avenir" LabEx PALM (ANR-10-LABX-0039-PALM) is also gratefully acknowledged.
\end{acknowledgments}

\end{document}